# A LIGHT TRANSFORMER FOR SPEECH-TO-INTENT APPLICATIONS

*Pu Wang, Hugo Van hamme*

KU Leuven, Department of Electrical Engineering-ESAT
Kasteelpark Arenberg 10, Bus 2441, B-3001 Leuven Belgium
{pu.wang, hugo.vanhamme}@esat.kuleuven.be

## ABSTRACT

Spoken language understanding (SLU) systems can make life more agreeable, safer (e.g. in a car) or can increase the independence of physically challenged users. However, due to the many sources of variation in speech, a well-trained system is hard to transfer to other conditions like a different language or to speech impaired users. A remedy is to design a user-taught SLU system that can learn fully from scratch from users' demonstrations, which in turn requires that the system's model quickly converges after only a few training samples. In this paper, we propose a light transformer structure by using a simplified relative position encoding with the goal to reduce the model size and improve efficiency. The light transformer works as an alternative speech encoder for an existing user-taught multitask SLU system. Experimental results on three datasets with challenging speech conditions prove our approach outperforms the existed system and other state-of-art models with half of the original model size and training time.

## 1. INTRODUCTION

The growing acceptance of voice assistants such as Siri and Google Home lead to a regained interest in spoken language understanding (SLU) systems for voice command-and-control (C&C) applications. People, especially elderly and physically challenged people, could benefit a lot by uttering instructions through a voice interface to get assistance from e.g. a smart home system, domestic appliances or a robot. It leads to fairly simple commands like "turn the heat up in the bathroom". A conventional SLU system has a pipeline structure with two separate modules: automatic speech recognition (ASR) maps speech input to an intermediate textual representation and a natural language understanding (NLU) system decodes a semantic representation from the text [1]. To eliminate errors introduced by separate training, the pipeline structure is replaced by an end-to-end speech-to-intent (S2I) system which directly maps the speech features to predefined task like {*action: "increase", object: "temperature", where: "bathroom"*} without explicit text scripts [2, 3].

Typically, end-to-end S2I systems are constructed by deep learning approaches, which automatically learn projections between speech inputs and labeled outputs from a large amount of training data [4]. The challenge is the diverse nature of speech, as multiple users have different linguistic habits. A system trained on a specific domain is hard to transfer to other working conditions, such as a different language, non-standard grammar or thick dialects [5, 6].

A remedy presented in [5, 6] is to design a user-taught SLU system that fully trains on demonstrations from users. Since it allows users to give their own speech command and corresponding task label for training, the above restrictions do not apply. The disadvantage is it needs training samples and teaching time from users. To minimize the users' efforts, [5, 6] are both built with an encoder-decoder structure with a capsule network, which is believed to be a less data-hungry model [7], as the decoder. These systems utilize a bi-directional GRU as the speech encoder. The inherently sequential computation enlarges the scale of these two systems and slow down the training [8].

An alternative approach to capture the complex relations between spectral features/phones/words in speech is the transformer structure, which is based on efficient parallel computation [9]. In this paper, we will utilize a variant of transformers to replace the GRU in the aforementioned system as the speech encoder and investigate whether comparative performances can be obtained with less training effort. The original SLU system from [5, 6] will serve as the baseline.

The major issue for transformers is their insensitivity to the sequence order as it simultaneously attends to each frame of input sequence by the self-attention mechanism. Considering the position information is indispensable for speech processing, attention-based models have therefore used additive position embeddings to encode the order in a sequence [9]. However, because position embeddings are added to the features, the model will have to learn in which subspace relevant data variation occurs and in which subspace position is represented. This may be successful when large amounts of training data are available. In our context, data is scarce since the user needs to perform demonstrations to teach the system. In [10], a relative position encoding is proposed which separates the content and location subspace. Since transformers with relative position encoding have also been proven better than with absolute position [10-12], the method in [10] is our first choice. However, it introduces several additional location-based trainable parameters which inevitably enlarge the model scale especially for high-dimensional features. A remedy is to replace the addition with concatenation, which naturally pre-defines a content-subspace and a location-subspace. With constraints imposed on the trainable parameters, the model's data requirement will be low. Also, we argue a low dimensional position representation should suffice, which reduces the model size further.

So, in this paper we will propose a light version of transformers which includes a low-dimensional (light) relative position encoding matrix. This model will serve as the speech encoder to help the existing user-taught SLU systems gain a more compact structure and get a more efficient performance by reducing the large number of trainable parameters. Besides its 'light' size of the model scale, the light transformer is also 'light' for memory consumption via a 4-fold down-sampling of the input sequence by using convolutional layers and sharing all cross-layer parameters. The contribution of this paper is hence the encoder design that requires fewer training samples, i.e. demonstration time from its users. In this work, an important evaluation criterion will hence be the number of training samples required for a given accuracy.

The original transformer and the proposed light version will first be explained in section 2. Some related memory-saving strategies along with the whole structure of the user-taught SLU system will also be given in this section. Section 3 discusses the specific experimental setting for evaluation, and the corresponding results will be presented in section 4. In section 5 we will conclude our work.

## 2. MODEL

### 2.1. The basic transformer with relative position encoding

The transformer is a layer-stacked model with each layer consisting of two sub-layers: the multi-head self-attention layer:

$$MultiHead(x) = W_{out} Concat(Attn^1, Attn^2, \cdots Attn^N)$$

and the fully connected feed-forward layer:

$$FFN(x) = W_2 \max(0, W_1 x + b_1) + b_2.$$

The output of the multi-head attention is a linear projection with the weight matrix $W_{out}$ of a concatenated result from multiple scaled dot-product self-attention mechanism, which is denoted as $Attn^n$.

In each attention head, the feature representation of the sequential data is first linearly transformed into a sequence of Keys, Values and Queries. The feature representation in the next layer is built as a non-linear mapping of a weighted average of the Values. The weight of each Value is determined by the similarity between the Key and the Query, as measured by a dot-product. To account for order information, a position encoding is typically added to the content embedding at the very beginning. The general mathematical expression is as equation (1):

$$Attn_i = softmax \frac{(Kx_j + Kp_j)^T (Qx_i + Qp_i)}{\sqrt{d_k}} (Vx_j + Vp_j) \quad (1)$$

where $x$ is the feature representation or so-called content embedding and $p$ is the position embedding. $Q, K, V \in \mathbb{R}^{d_{model} \times d_k}$ are weight matrices of the Query, the Key, and the Value respectively. $d_k$ is the dimension of the content embedding $x$ and position embedding $p$. $d_{model}$ is the dimension of the transformer model.

The attention weights computed as equation (1) only considers the absolute position information, not the connection of arbitrary relative location. But the later one makes more sense in practice, especially when the same words occur in one sentence more than once. To address this shortage, Dai [10] modified equation (1) to include the relative position information.

$$Attn_i = softmax \frac{(Kx_j)^T(Qx_i) + (K_p p_{i-j})^T(Qx_i) + (Kx_j)^T u + (K_p p_{i-j})^T v}{\sqrt{d_k}} (Vx_j) \quad (2)$$

where $p_{i-j} \in \mathbb{R}^{d_p}$ is the position shift between the Query and the Key. The learnable vector $u, v \in \mathbb{R}^{d_p}$ are introduced to replace the $Qp_i$ term of equation (1).

The relative position embedding has been proved to perform much better [10-12]. However, it adds three extra location-based learnable parameters: $u, v \in \mathbb{R}^{d_p}$, $K_p \in \mathbb{R}^{d_{model} \times d_p}$.

### 2.2. The light transformer with low-dimensional relative position encoding

The most popular position embedding is defined as the following:

$$PE(t, 2i) = sin(t/T^{2i/d_x})$$
$$PE(t, 2i + 1) = cos(t/T^{2i/d_x})$$

where $t$ is the position and $i$ is the dimension. $T$ is the longest sequence length and $d_x$ is the dimension of the input sequence.

It is clear that the original position embedding depends on both time and feature dimension and leads to a large size of the $K_p, u$ and $v$ variables.

The alternative light choice is just encoding the essential time information: the position of each frame $t$, as shown below:

$$P(0, t) = cos(2\pi t/T)$$
$$P(1, t) = sin(2\pi t/T)$$

If $0 \leq t \leq T$, there will be no aliasing in the position and position $t$ is uniquely determined from the 2-dimensional vector $P$. But considering the differences between the $P$ will be small if the sequence length $T$ is large, following four terms are concatenated:

$$P(2,t) = cos(2\pi t/M_1)$$
$$P(3,t) = sin(2\pi t/M_1)$$
$$P(4,t) = cos(2\pi t/M_2)$$
$$P(5,t) = sin(2\pi t/M_2)$$

with $M_1$ and $M_2$ are small integers to provide word and phone resolution respectively.

This 6-dimensional position information is concatenated to the feature representations. As such, the inputs $x_{in}$ consist of two parts: content embedding and position embedding.

$$x_{in} = \begin{pmatrix} x \\ p \end{pmatrix}$$

After replacing additive position embedding with the concatenated one, the matrix $Q$ and $K$ in equation (1) are reshaped to the block-diagonal structures.

$$Q = \begin{bmatrix} Q_c & 0 \\ 0 & Q_p \end{bmatrix}, K = \begin{bmatrix} K_c & 0 \\ 0 & K_p \end{bmatrix}$$

where $Q_c$ and $K_c$ are content-related weight matrixes, $Q_p$ and $K_p$ are position-related weights. Notice that the position embedding now has dimension $d_p = 6$, which differs from the dimension of the content embedding $d_k$. The equation (1) is therefore modified to

$$Attn_i = softmax\left(\frac{(K_c x_j)^T(Q_c x_i)}{\sqrt{d_k}} + \frac{(K_p p_j)^T(Q_p p_i)}{\sqrt{d_p}}\right)(V x_j) \quad (3)$$

Here, the Value transformation takes only the content embedding. So, attention weight is position sensitive, but the content vector is not. To include the position information in each transformer layer, the 6-dimensional position embedding is introduced to every layer, as well as the final outputs.

Referring to [10], the equation (3) is further revised to consider arbitrary relative relations between Query and Key as follows:

$$Attn_i = softmax\left(\frac{(K_c x_j)^T(Q_c x_i)}{\sqrt{d_k}} + \frac{(K_p p_{i-j})^T u}{\sqrt{d_p}}\right)(V x_j) \quad (4)$$

Compared with the equation (2), we only introduce two trainable parameters $K_p$ and $u$. Moreover, their 6-dimensional scale is much smaller than the original ones.

Besides the modified position encoding and corresponding attention computation, we preserve the multi-head structure, position-wise feed-forward layer and residual connection of the original transformers.

### 2.3. Memory issue

To discuss the memory usage of the transformer, let us focus on the equation (4). Considering the input data has shape [batch size, dim$_{inputs}$, length], assuming the Query, Key and Value have the same dimension dim$_{QKV}$, for multi-head attention with N heads, the vector $Q_c x$, $K_c x$ and $Vx$ all have the shape [batch size, N, length, dim$_{QKV}$]. The main issue is the term $(K_c x_j)^T(Q_c x_i)$ [13], which, if unrestricted, has the shape [batch size, N, length, length], its memory consumption grows quadratically with the sequence length and linearly with the number of heads.

Transformers with more attention heads no doubt perform better, but easily hit the memory limitations. Therefore, we try to use less attention heads but with larger inner depth of the feed-forward layer to compensate for such loss. It makes sense as the feed-forward layer is used to project the attention outputs potentially giving it a richer representation. Because the wider feed-forward layers also account for a larger fraction of memory use, we would share feed-forward layer as well as attention parameters across layers [14]. To reduce sequence length, we use two 2-dimensional convolution layers with kernel size (5,5) at the very beginning to do 4-fold down-sampling in time.

### 2.4. Multitask user-taught SLU system

The whole structure of the user-taught SLU system is shown in Figure 1. Its general encoder-decoder concept is explained in detail in [5], while [6] extends it to the multitask application with speaker identification function, which lets it works for multiple users. Since the code of [5] is available[1], we first follow the instruction in [6] to reconstruct it to the multitask system. The main change of this paper is replacing the GRU with the light transformer encoder.

The light transformer consists of 4 layers. Every layer has an 8-head parallel attention layer with 64 dimensional heads and a feed-forward layer of dimensionality 2048. $M_1$ and $M_2$ of position matrix are 4 and 2. To further speed up the training, self-attentions were restricted to considering only a neighborhood of size 5 centered around the respective output position. A residual connection and layer normalization are added around each sub-layer to ensure a stable backpropagation and gradient.

The filterbank inputs are first processed by the light transformer encoder to yield the high-level representation. The "distributor & attention" mechanism then filters out the unimportant segments of the encoded features. Finally, the capsule network decoder yields the task and speaker information from the output capsule. The detailed description of capsule decoder and distributor & attention could be found on [5, 6]. To conduct the comparison, the hyper parameters of the remaining decoder are chosen as in [6] without alteration.

---

[1] https://github.com/vrenkens/assist

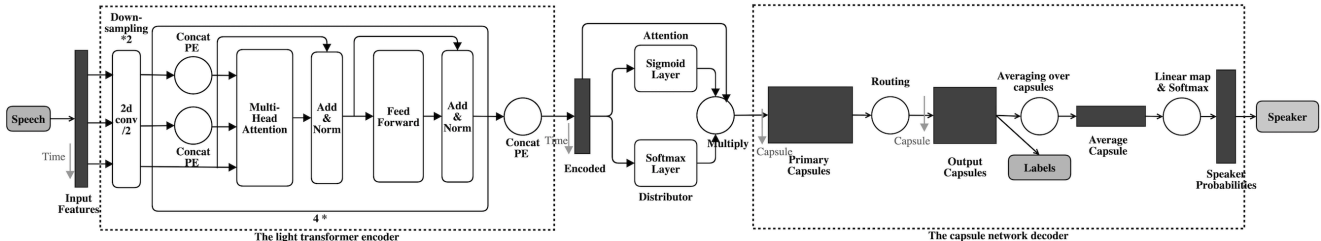

Figure 1: Structure of the multitask user-taught SLU system

## 3. EXPERIMENTS

### 3.1. Dataset

The model will be tested on three public datasets. Each dataset simulates one challenged spoken situation, including different language speakers, varied command phrasings, and impaired speech.

**Grabo** [15] records commands to robots, like "drive quickly to the right". It is a dataset with commands that have a similar structure with limited linguistic variation. It contains a total of 6000 records for 33 task labels about positions, movement speeds and actions from 10 Dutch speakers and 1 English speaker.

**FluentSpeech Commands** [16] is an English smart-home setting for virtual assistant purposes, including commands like "turn up the heat in the kitchen". With 30043 utterances this dataset is larger and also more challenging. There are 248 different phrasings for 31 distinct intents from 97 speakers.

**Domotica-4** [17] contains utterances for home automation tasks, like "Front door Open", from 7 dysarthric Dutch speakers. There are 22 different output labels.

### 3.2. Experimental Setup

For the Grabo and Domotica-4 dataset, we mainly compare the learning curves of the F1-measure for intent label classification and percentage of correctly decoded speakers. The learning curve records the model's performance on an increasing amount of training data and tests on all remaining data using 5-fold cross-validation. The utterances from all speakers are randomly shuffled beforehand and then all data is divided into 150 blocks [5]. Starting from one block, the model is trained on an increasing number of blocks and tested on all remaining blocks. This way a learning curve is created [6].

For FluentSpeech Commands, besides the learning curve, we also simulate the insufficient training data situation for intent classification task. The full dataset is divided into train set (23132 utterances), valid set (3118 utterances) and test set (3793 utterances). We randomly select 10%, 30%, 60% and 100% data from the train set, test on the test set to get intent accuracy results. The accuracy matrix is defined in [18].

### 3.3. Hyper parameters

We denote the proposed SLU system here as the "light transformer-capsule" model. It is trained with the Adam optimizer with $\beta_1 = 0.9, \beta_2 = 0.98$ and $\epsilon = 10^{-9}$. We also varied the learning rate using the warmup setup as suggested by Vaswani [9]. The final model is constructed by averaging the model parameters stored at the last 10 training steps. To regularize during training, we apply dropout with a rate of $P_{drop} = 0.1$ to each sub-layer, including the content and position embeddings.

### 3.4. Baseline

**GRU-capsule model:** The original multitask SLU system presented in [6] serves as the baseline for these three datasets. We denote it as the "GRU-capsule" model. It includes a 2-layer bidirectional GRU with 256 units in each direction as the encoder, the remaining hyper parameters of the decoder are the same as the light transformer-capsule model.

**SincNet/DFSMN-transformer:** An end-to-end approach proposed by [19] which works on FluentSpeech Commands dataset. It utilizes SincNet and DFSMN as the acoustic model component and a transformer encoder as the SLU component.

**SOTA transformer:** A state-of-art end-to-end SLU model fully based on transformers [20] which was evaluated on the FluentSpeech Commands dataset. We compare to the best results in [20] from its classification-based model.

**SOTA RNN:** A state-of-art bidirectional RNN encoder based end-to-end model presented in [18] designed for the FluentSpeech Commands dataset. Since none of the methods discussed here uses pre-training (its applicability is questionable given the multilingual setting of Grabo and the dysarthric speech in Domotica-4) we compare to the results in [18] without pre-training.

## 4. RESULTS

### 4.1. Grabo dataset

In Figure 2, we present compared results on the Grabo dataset between the light transformer-capsule and the GRU-capsule. The count of trainable parameters amount is also summarized in Table 1 for model scale comparison.

Table 1: Comparison of the number of parameters among the test multitask SLU systems

| Model | # of parameters |
|---|---|
| light transformer-capsule | 1.31M |
| GRU-capsule | 2.36M |
| base abs-transformer-capsule | 1.31M |
| base rel-transformer-capsule | 1.39M |

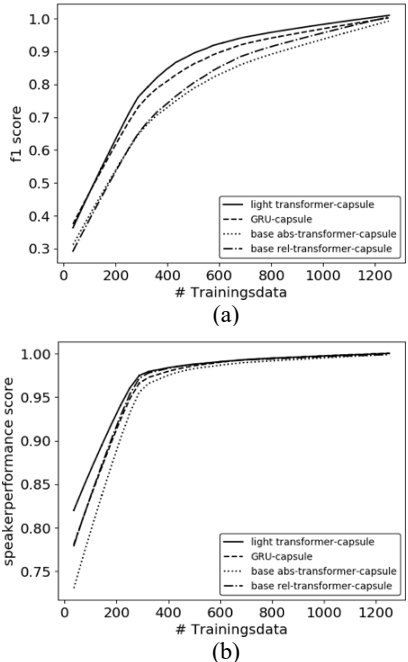

Figure 2: Learning curves of (a) F1-score of intent mapping; (b) percentage of correct speaker identification on the Grabo dataset among the test multitask SLU systems.

It is remarkable that the light transformer-capsule gets comparable results with around the half model size of the original GRU-capsule model. Such reduction on the number of trainable parameters helps the model to converge quickly. We will further quantitatively discuss the efficiency of compact structure on the FluentSpeech Commands dataset. Moreover, due to the lighter structure of the light transformer, the system is easier trained on smaller data size and gains better performance.

We also include the results from the basic transformer with both additive absolute and relative position encoding, which are denoted as the "base abs-transformer-capsule" and the "base rel-transformer-capsule" respectively, to further investigate the light transformer's ability. The base abs-transformer/rel-transformer all utilize the same hyper parameters and memory-saving strategies.

As shown in Table 1 the light transformer has almost the same size as the base transformer with additive absolute position, while being more compact than the additive relative position model. From Figure 2, it is clear that the light transformer with concatenated position information significantly outperforms the other two additive ones especially with fewer training samples, which confirms that the concatenation operation used by the light transformer brings a better parametrization for user-taught SLU. Moreover, the performance gains w.r.t. standard approaches observed on this speech processing task creates the hope other SLU or NLP tasks could benefit from light transformers. Meanwhile, the relative position embedding is better than the absolute one, which motivates our choice in the light transformer.

### 4.2. FluentSpeech Commands dataset

Following the experimental setting in [18, 19], we simulate an insufficient training situation by randomly choosing 10%, 30%, 60% and 100% data from the train set. Table 2 summarizes intent accuracy results on the FluentSpeech test set. Since the light transformer-capsule does not get prior-knowledge from pre-training, the two state-of-art model [18, 19] serve for comparison here also work without pre-training. From Table 2, given different sizes of the train sets, the light transformer-capsule outperforms all other state-of-art models. Also, compared with the full base transformer structure from [20], the proposed light transformer outperforms the former by 1.2% absolute which further demonstrates the proposed structure's capacity for sequence modeling.

Table 2: Comparison of intent accuracy among different approaches on the FluentSpecch Commands dataset

| # of training data | 10% | 30% | 60% | 100% |
|---|---|---|---|---|
|  | 2.3k | 6.9k | 13.8k | 23.1k |
| light transformer - capsule | 91.8% | 96.7% | 98.3% | 98.8% |
| GRU-capsule | 81.5% | 94.9% | 96.7% | 98.1% |
| SincNet/DFSMN-transformer | - | 92.9% | 97.0% | 98.1% |
| SOTA transformer | - | - | - | 97.6% |
| SOTA RNN | 88.9% | - | - | 96.6% |

Table 3: Comparison of training time among different approaches on the FluentSpecch Commands dataset

|  | light transformer-capsule (# of hours) | GRU-capsule (# of hours) |
|---|---|---|
| 10% train set | ca 0.5 | 1.5 |
| 30% train set | ca 1.5 | ca 4 |
| 60% train set | ca 3 | ca 7.8 |
| full train set | ca 5.5 | ca 13 |

To further discuss the efficiency of the light transformer, Table 3 records the training time (until model convergence) of the light transformer and the baseline GRU. The convergence curve on the validation set with 10% train data is also given in Figure 3. Apparently, the light transformer converges faster and saves more than 50% training time.

Figure 4 shows the learning curves of intent and speaker classification. In both tasks, the light transformer shows improved performance (up to 10%) when it comes to limited training data, which confirms that the light transformer better meets the requirements for a user-taught SLU system.

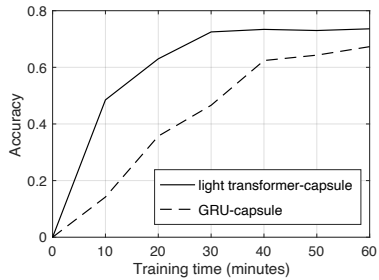

Figure 3: Intent accuracy on the validate set over time for models training on 10% train data.

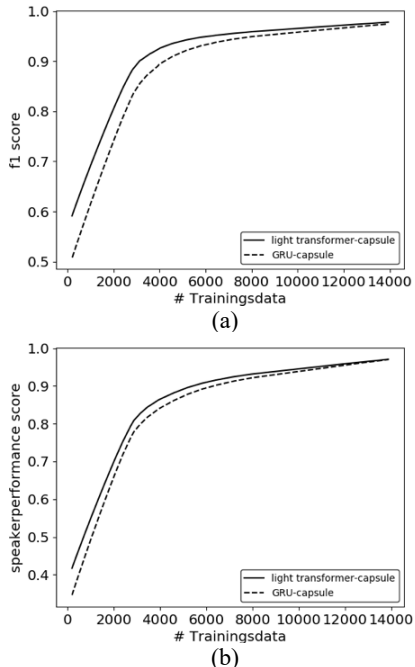

Figure 4: Learning curves of (a) F1-score of intent mapping; (b) percentage of correct speaker identification on the FluentSpeech Commands among the test multitask SLU systems.

### 4.3. Domotica-4 dataset

The results on the Domotica-4 dataset are shown in Figure 5. Although the light transformer-capsule still outperforms the GRU-capsule in general, we notice the intent classification of the former is slightly worse than the baseline for less than 100 training samples.

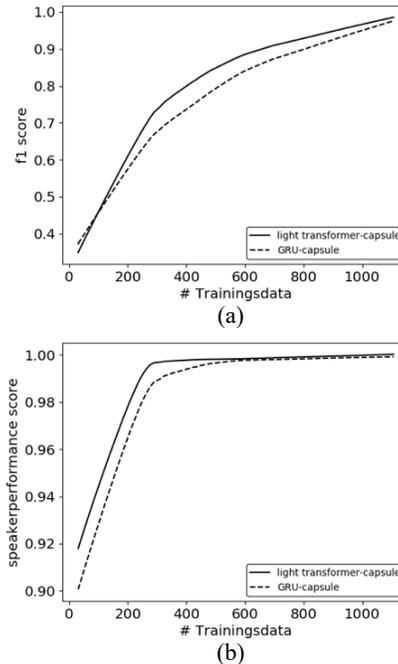

Figure. 5 Learning curves of (a) F1-score of intent mapping; (b) percentage of correct speaker identification on the Domotica-4 among the test multitask SLU systems.

## 5. CONCLUSION

In this paper, we proposed a light version of the transformer encoder by concatenating a low-dimensional relative position embedding. Its memory consumption is further saved by introducing several strategies. Mathematically, the proposed structure has fewer trainable parameters and less memory burden compared with the basic relative position transformer.

The light transformer encoder is proposed for user-taught direct S2I mapping, which requires the presented system to possess a compact and highly efficient structure without compromising the performance. The proposed model then replaces the GRU encoder of an existing GRU-capsule SLU model to form a light transformer-capsule network system.

Experimental results on three public datasets verify that the light transformer-capsule model significantly scales down the model size to almost half and speeds up the convergence process to save more than 50% training time. Moreover, it outperforms the original GRU-capsule and other state-of-art models including the basic transformers with additive position embedding on both intent mapping and speaker identification tasks.

## 6. ACKNOWLEDGEMENTS

The research was supported by the program of China Scholarship Council No. 201906090275, KUL grant CELSA/18/027 and the Flemish Government under "Onderzoeksprogramma AI Vlaanderen".